# Real-time analysis of mechanical and electrical resonances with open source sound card software

**G Makan[1], K Kopasz[2] and Z Gingl[1]**

[1]Department of Technical Informatics, University of Szeged, Árpád tér 2, 6720 Szeged, Hungary

[2]Department of Optics and Quantum Electronics, University of Szeged, Dóm tér 9, 6720 Szeged, Hungary

E-mail: makan@inf.u-szeged.hu

**Abstract.** We present an easily reproducible, open-source, sound card based experimental set-up to support transfer function measurement. Our system is able to visualize signals of mechanical and electrical resonances and their spectra in real time. We give a brief description of the system, and show some examples of electrical and mechanical resonance experiments that are supported by the system. The theoretical background, experimental set-up, component selection and digital signal processing are all discussed, and more detailed information (building instructions, software download) is provided on a dedicated web page (http://www.noise.inf.u-szeged.hu/edudev/RealTimeAnalysisOfResonances/) The experimental set-up can support undergraduate and graduate education of students of physics, physics education and engineering by means of experimental demonstrations and laboratory exercises. The very low cost, high efficiency and transparent system provides a scalable experimental environment that can be easily built in several instances.

## 1. Introduction

Modernization and innovation is a continuous challenge in physics education. Laboratory budgets are usually not large enough to buy expensive devices; therefore cheap but useful solutions are much in demand to support experimental education. Increasing the motivation and participation of students is another necessary aspect of innovation and involving computer applications in education is a good way to achieve this goal. There are various animations and simulations [1] that could be presented in physics education to demonstrate a lot of physical phenomena, but they cannot replace the real experiments. Nevertheless, computers can be integrated as part of this kind of demonstrations. Using the so-called virtual instrumentation technique considerable part of the instrument is represented by software, although real measurement is performed. Sensors, electrical circuits and data converters can be used to translate the real physical signals into digital signals represented by numbers that can be efficiently handled and visualized by computers in real time. Using this method, students can gain direct experimental knowledge and, at the same time, learn to apply modern methods and tools to improve their own important skills concerning instrumentation, data analysis and programming.

A rather important topic of physics education is the demonstration of mechanical and electrical resonances. Although lecturers usually talk about RLC circuits, they cannot easily carry out experiments with traditional devices (e.g. with expensive oscilloscopes). However, because of the low abstraction level of the students, experimenting is a very helpful tool in the teaching of this part of the curriculum, especially at the undergraduate level.

As it has been known for many years [2-4], oscilloscopes can be replaced with sound cards with certain limitations and many kinds of signal processing (including spectral analysis) can be implemented. The input of the RLC circuit can be driven by the output of a PC sound card, while the signals at the output of the circuit can be detected by the input of the sound card. During our work we studied several sound card based oscilloscopes and spectrum analysers that are publicly available on the web [5-14]. We especially looked for free, open-source software that allows performing real-time spectral analysis and transfer function measurement in a reliable way. Although the available applications can help in many cases, real-time measurements of system transfer functions are not supported at this level required for high quality education.

Keeping these in mind, we have developed a very low cost sound card based measuring and experimenting system especially dedicated to support practical investigation of resonances in electric and mechanic systems at various levels of undergraduate and graduate education. We have built an application in the LabVIEW environment, which has a powerful and easy-to-use graphical programming language. Based on our experience, undergraduate and graduate students, PhD students and physics teachers – even those ones who are not too proficient in programming – are able to learn to use it very quickly. LabVIEW is extensively used in research and industry and has become a standard tool in computerized instrumentation worldwide. The source code of our software is publicly available on our web page [15], so it can be downloaded, can studied and modified as required.

Our measurement set-up can be used for real-time system analysis due to the fast update rate capability of 10 measurements per second, and supports studying periodic and transient signals in a cheap yet accurate and reliable way. It allows lecturers to demonstrate almost any kind of experiments related to the detailed analysis of resonances. Another advantage is that it can be used in a laboratory environment to carry out exercises common to all students, while creative students are able to build their own set-up and do or further develop experiments at home or as a part of an experimental homework.

In the next section we briefly describe the system and the types of continuous and transient excitations that can be used to measure the system transfer function. We also show how the similarities between mechanical and electrical systems could contribute to the knowledge about the physical phenomena, sensor applications, data conversion and digital signal analysis.

## 2. Experiments

Sound card outputs and inputs can support signal generation and data acquisition systems using the built in high-resolution data converters – digital-to-analogue (D/A) and analogue-to-digital (A/D) converters. In order to protect the built-in sound card of the PC from the higher loads and possible excessive transients one can use an external sound card connected via one of the USB ports available on any computer today. Important to note that this also makes easier to carry out the measurements reliably on any computer since the sound cards can be different in different computers. An average sound card is able to generate and measure signals from 20 Hz to 20,000 Hz in a small, about 2.5 V voltage range (-1.25V to 1.25V) in most cases. We have used the C-media CM119A type external sound card [16] which is small, cheap and of sufficient quality for the purpose of the experiments.

Most of the sound card based measuring systems only produce qualitative results regarding amplitude accuracy. However, with our set-up, it is possible to measure the relative amplitude and the absolute frequency of a signal precisely. We also provide a simple calibration procedure to enhance accuracy. Indeed, since the A/D converter of the sound card has a 16-bit resolution, one can achieve a significantly higher resolution than with a sound card in a normal oscilloscope, typically working with an 8-bit resolution. Using the TINA circuit simulation software [17] we have analysed the model of the RLC circuit (see figure 1) and found that if the resonance frequency of the circuit is in the frequency range of the sound card, then the impedances belong to the input and the output of the sound card do not affect the measurements considerably. According to equation (1), if the nominal values are L = 330 mH and C = 4.7 mF, then the calculated resonance frequency (*f*) is 4040 Hz, which agrees with the result obtained from the TINA simulation.

$$f = \frac{1}{2\pi\sqrt{LC}} \tag{1}$$

From the real measurements we got 4050 Hz for *f* (see later), which is in a very good agreement considering the typical tolerance limits of the components and the frequency resolution of the measurement. The frequency characteristics of the CM119A sound card are satisfying, however although the time bases of the output and the microphone are found to be slightly different, so the measured output signal drifts slowly over time. However, it does not cause any error in the case of our measurements since the input and output signals are measured simultaneously and both the amplitude and phase of the transfer function can be measured precisely.

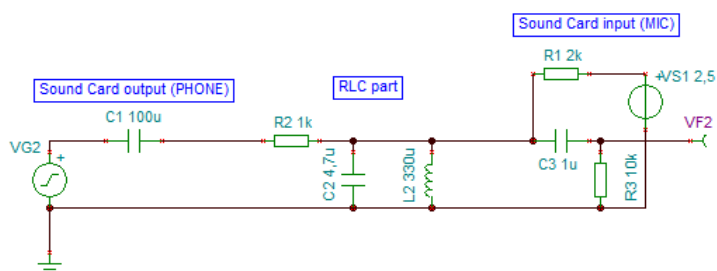

**Figure 1.** The model of the parallel RLC circuit created with TINA. The aim was to investigate how the structure of the input and output of the sound card affect the resonance frequency measurements.

*2.1. Electric resonances*

Our software is able to generate four different types of excitations that can be used to measure the transfer function of the investigated system. We can excite the electrical and mechanical systems with single frequency signals that are swept over a frequency range; with impulses; with chirp-pattern signals and with noise (so called "periodic random noise"). The first two methods can be considered as classical, since they are closely related to the theoretical introduction to the linear system transfer function analysis: response to a single sinusoidal can be used to see the amplitude and phase changes and in the case of a sharp impulse excitation – whose spectrum is flat, the initial phases are zero – the output signal is identical to the transfer function of the system. However from a technical and instrumentation point of view – that is rather important in education of physics as well – using the chirp pattern or noise with the associated digital signal processing is a modern and efficient way of measurement. We used a certain kind of white noise, the periodic random noise (PRN) that is a sum of sine waves of different frequencies, identical amplitude and random initial phase. These latter two signals provide well distributed energy in the desired time and frequency range, so the signal-to-noise ratio is significantly better than for an impulse-like excitation.

In order to determine the transfer function of a linear system first we consider the Fourier series expansion of a periodic signal *x*(*t*):

$$x(t) = \sum_{n=-\infty}^{\infty} c_n e^{i\omega_n t},$$

$$c_n = \frac{1}{T} \int_0^T x(t) e^{i\omega_n t} dt, \quad (n = 0, \pm 1, \pm 2, \dots)$$

$$\omega_n = n\frac{2\pi}{T},$$

(2)

where $c_n$ is the complex Fourier coefficient and $T$ is the period of the sine and the cosine wave. In the case of a single input-output system the response to the impulse excitation is the so called impulse response. If we know the impulse response of a system, we can give the output signal for any input signals: we simply have to take the convolution of the input signal and the impulse response. In the most cases, the frequency response of a system is given instead the impulse response, what we can get from the Fourier transform of equation (3):

$$x(t) * h(t) = y(t), \tag{3}$$

$$X(f) \cdot H(f) = Y(f), \tag{4}$$

$$H(f) = \frac{Y(f)}{X(f)}, \tag{5}$$

where $x(t)$ and $y(t)$ are the input and output signals, respectively, and $h(t)$ is the impulse response of the system. $X(f)$ and $Y(f)$ are the Fourier transformed functions of the input and the output signals, respectively, while $H(f)$ is the frequency response function [18]. In the case of finite and sampled functions one should use the Discrete Fourier Transform (DFT) to transform them in frequency space.

$$X_k = \sum_{n=0}^{N-1} x_n \cdot e^{-i2\pi kn/N}. \tag{6}$$

In the case of measured, time-depending signals, $X_k$ is the magnitude of the given frequency, $x_n$ is the value of the n-th sample, and N is the number of samples.

*2.1.1. Frequency response of an RLC circuit excited with* periodic random noise

In our experiments the most spectacular one is the frequency response of the parallel RLC circuit; but on the prototype board (figure 3) one can build different type of configurations as well. In the case of the parallel connection we used an RLC circuit at the resonance frequency 4040 Hz (R=200 ohm, L=330 uH, C=4.7 uF).
Since the periodic random noise is generated using random numbers (initial phases) in the frequency domain, it has advantages over using white noise that can be represented by a sequence of random, uncorrelated amplitude values generated in the time domain. The amplitude of the sinusoidal components is identical, therefore no averaging is required, and the desired frequency range can be covered. On the other hand, by further optimizing the random phases, the crest-factor of the signal can be minimized that allows higher excitation power and enhanced signal-to-noise ratio.
Exciting with repeated signals supports real-time spectral analysis [19]. The continuously repeated periodic random noise sample with length of 0.1s allows doing measurements with 10 Hz frequency resolution and update rate of 10 measurements per second. The main adjustable parameters of the PRN are the frequency range, the number of frequency components and the frequency scaling of the generated components. At the display, one can easily set the scaling of the axes, as well as the rate of averaging and the windowing.

The frequency response of the parallel RLC circuit in the case of PRN excitation is shown on figure 2. The measured resonance frequency is 4050 Hz note that the frequency and magnitude scales are logarithmic. We used this configuration in every case of experiments are listed below.

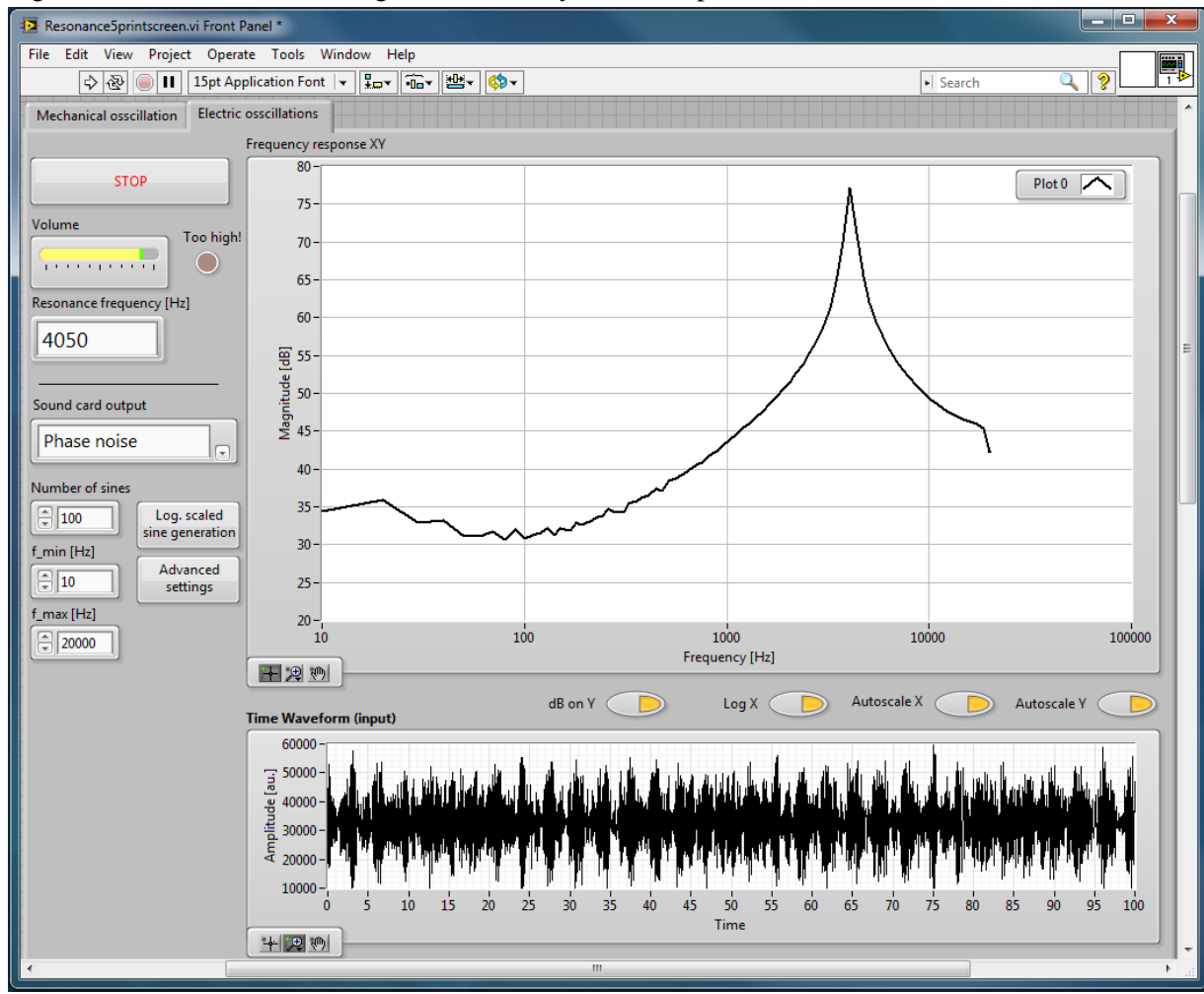

**Figure 2.** The graphical user interface of our software. The upper plot shows the frequency response of the RLC circuit. On the lower plot the input signal of the RLC circuit is shown for the case of periodic random noise excitation. There are controls and indicators on the left side to adjust measurement parameters.

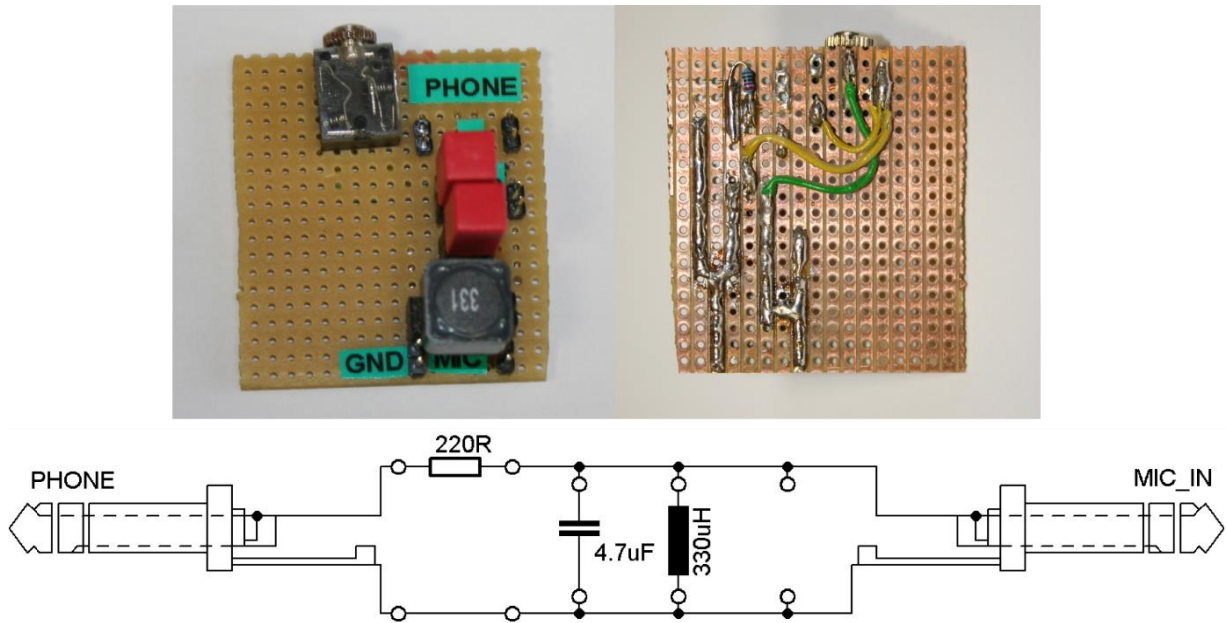

**Figure 3.** The parallel RLC circuit built on the perforated prototype board and its schematic diagram. The components are connected with jumpers to allow changing between the different configurations [20].

The quality of the cheap external sound cards is usually lower than the built in PC sound cards. With our software it is possible to correct the roughness of the frequency response of the sound card with a calibration file. To create this file, one has to connect directly the output of the sound card to its input and a calibration measurement should be performed. Figure 4 shows an example uncalibrated and calibrated frequency response of the sound card.

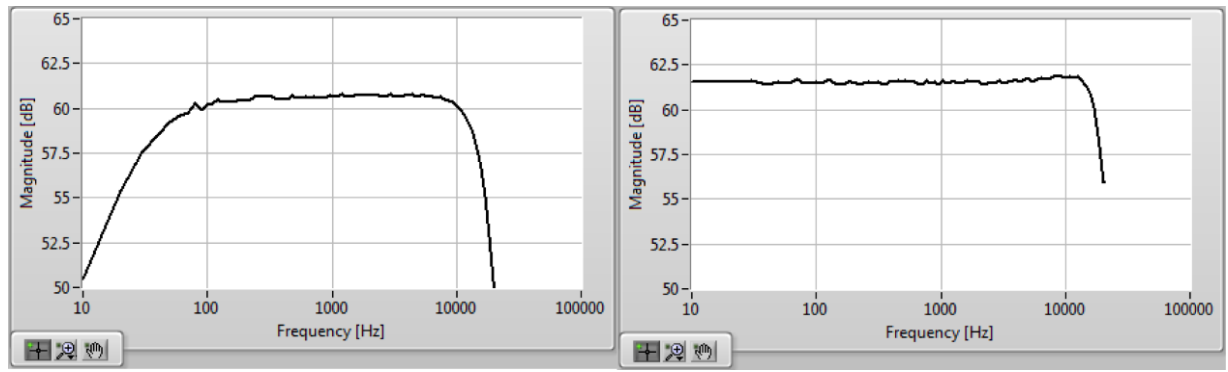

**Figure 4.** Calibration of the frequency response of the external sound card. The original response is shown on the left side, while the calibrated frequency response is plotted on the right side.

*2.1.2.Impulse excitation technique*

Impulse excitation technique is a frequently used method to obtain the impulse response function in the time domain directly. In this case a short voltage peak is applied on the input of the circuit under investigation. If the circuit is an RLC, the output signal will be a damped signal oscillating with the resonant frequency of the circuit (see the left hand side of figure 5). The automatic triggering function of the system analyser is useful, because we can observe a standing signal of the quite fast (10 ms long) damping, which depends on the value of the resistor (see the left hand side of figure 5). As it was mentioned already this kind of excitation signal is not practical for frequency domain analysis due to the low signal-to-noise ratio caused by localized energy in the time domain.

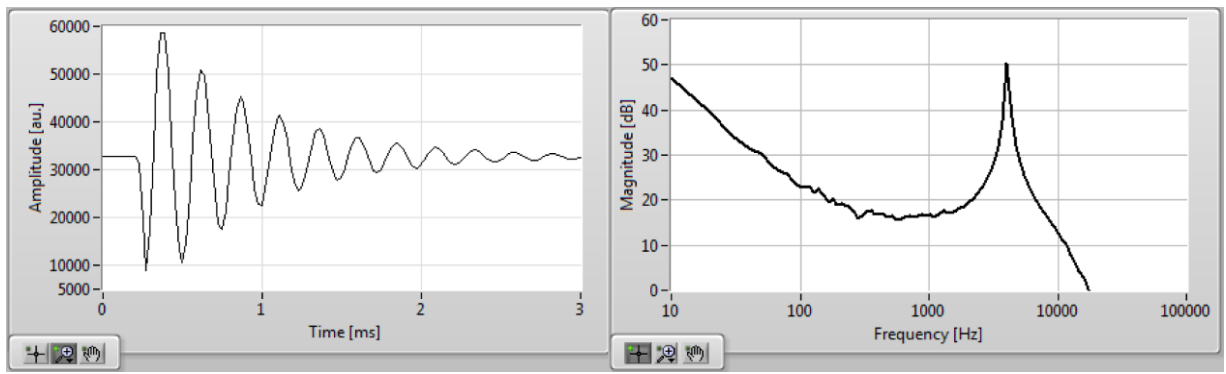

**Figure 5.** The impulse excitation technique allows the transient examination of a system. We can see the response signal on the left side and the frequency response on the right plot.

*2.1.3. Sweep excitation*

One of the most frequently used methods is frequency sweeping. This excitation contains only one sine wave whose frequency is growing in given steps, sweeping over an adjusted range (see figure 6). This method gives the best signal to noise ratio, but it usually requires a long time to measure the frequency response, therefore it cannot be used for a real-time measurement.

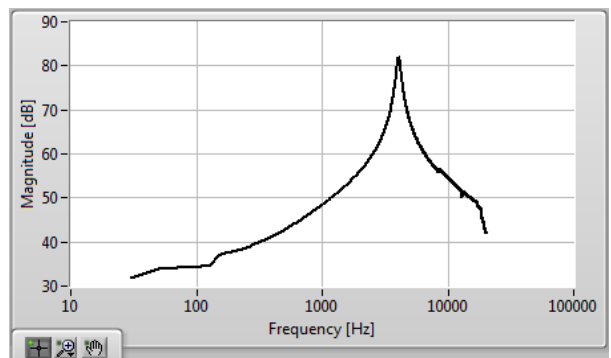

**Figure 6.** Frequency response of the measured system in the case of sweep excitation. The frequency range of the excitation and the step size are adjustable. Displaying is helped with a peak-hold function.

*2.1.4. Chirp excitation*

In a chirped signal sine waves with different frequencies are joined to each other in increasing frequency order. The length of the chirped signal is 0.1 s in our case, so this method allows efficient real-time spectral analysis. The power is distributed more or less equally in the desired frequency range although it can be generated in the time domain. Note that using the PRN is more straightforward since it is generated in the frequency domain and therefore it is easier for the students to understand how the spectrum of such a signal looks like. Figure 7 illustrates the use of a chirp signal for the transfer function measurement.

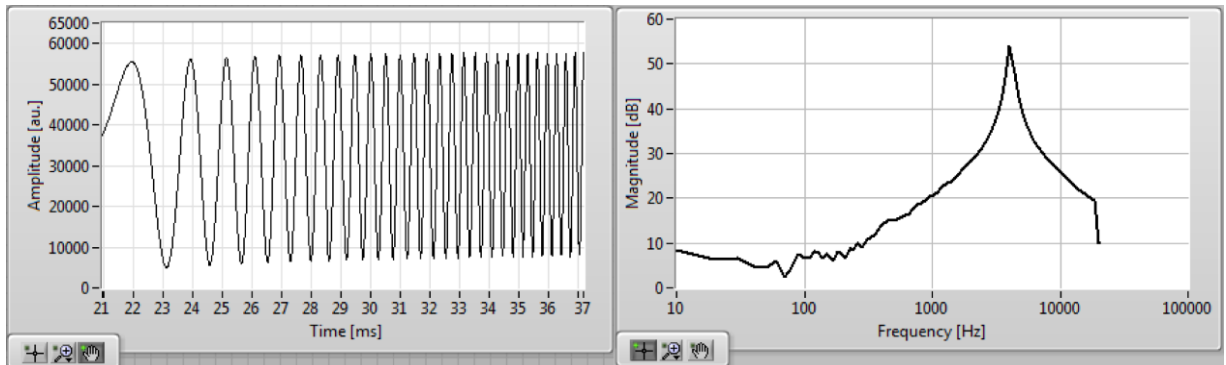

**Figure 7.** The time domain signal and the frequency response in the case of chirp excitation.

*2.1.5.2.1.5. Other examples for electrical experiments*

Real-time spectral analysis allows carrying out spectacular experiments to demonstrate resonance phenomena. In the simplest case one can connect an additional capacitor to the built-in one, which causes a jump of the frequency peak immediately. Using a potentiometer instead of the built-in resistor, one can adjust the quality factor and the magnitude of the resonance peak. Another interesting experiment can also be done by replacing the inductor with an air-core coil, and move an iron core into its center. In this case the induction of the cored coil will be growing continuously, while the iron core is moving inward; so the value of the resonance frequency will be decreasing (see figure 8). This can also be used to demonstrate the operation of inductive displacement sensors.

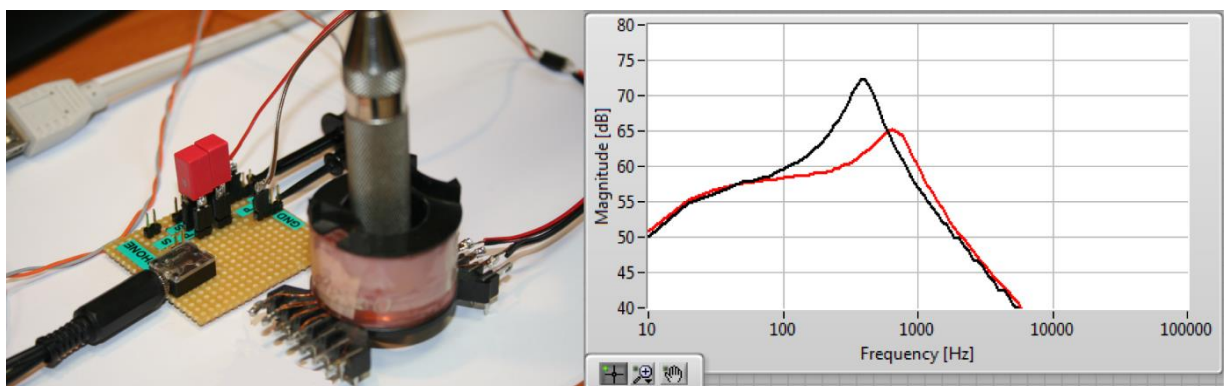

**Figure 8.** <u>Left</u>: An air-core coil connected parallel to the capacitor. <u>Right</u>: The resonance curve of the air-core coil (red line). If we use an iron core, the induction of the coil will be growing continuously, which results a decreasing resonance frequency (black line).

In addition, one can assemble different types of RC and RLC circuit configurations on a prototype board. For example, low pass, high-pass and band-pass filters can be realised and their transfer function measured. During electronics laboratory courses active filters and circuits can also be investigated.

*2.2. Mechanical resonances*

Resonance phenomena in mechanical systems are often considered first in introductory physics courses. Their response to excitations can also be described by frequency responses and transfer functions if they are operating in the linear limit, and they are good analogues of the electrical resonances.

Using adequate sensors, mechanical movements can also be examined with our set-up originally designed for studying electrical signals. There are examples for the use of sound card based measurements for this type of measurement in the literature. For example, Wild and Swan [21] examined spectra of musical instruments and steady waves in a one end opened pipe with frequency sweep excitation technique.

Here we present a measurement set-up which allows determining the frequency response of a tuning fork. We have used a sound card also for excitation and for data acquisition as in the cases discussed above. The excitation can be acoustic with a speaker that allows the use of many different kinds of excitation signals. Impulse-like mechanical excitation can be applied with a rubber hammer; it is analogous to the impulse excitation used in electrical experiments. The oscillation of the tuning fork was measured using a CNY70 [22] optical sensor which contains an infrared LED and a phototransistor. The infrared LED is driven by the 5V DC supply voltage came from a USB port, and a series resistor [20,23]. The phototransistor built in the sensor is connected directly to the microphone input of the sound card without any additional components [20]. The principle of the method is that the reflected infrared light amplitude is modulated by the oscillating tuning fork and the phototransistor translates it to an electrical signal that can be easily detected by the microphone input.

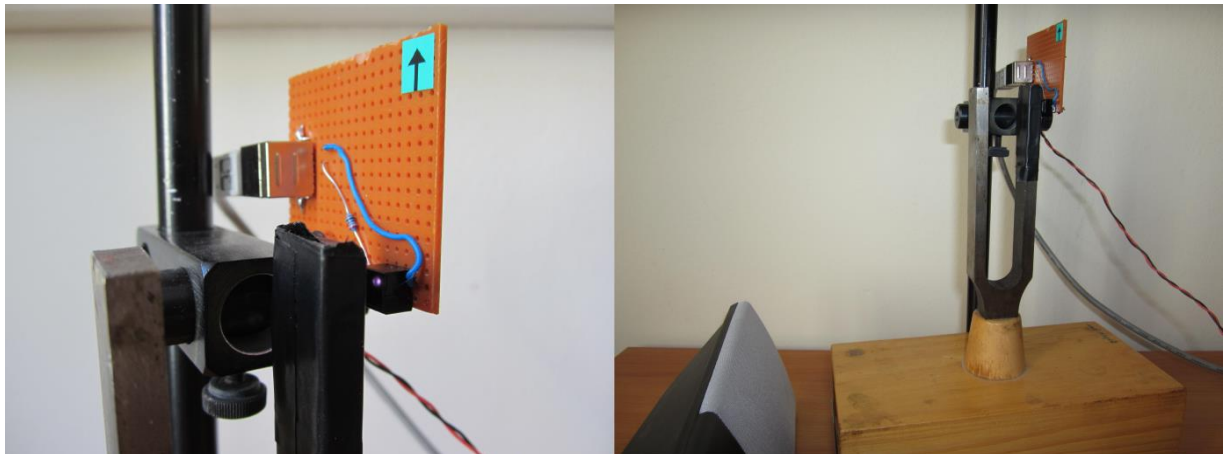

**Figure 9.** <u>Left</u>: the CNY70 type optical sensor and the USB connector built on a prototype board. <u>Right</u>: the tuning fork with a reflecting plate on its stem and the opposite placed prototype board.

*2.2.1. Acoustic excitation with a speaker*

In the case of a continuous acoustic excitation we can use exactly the same periodic random noise excitation technique like in electrical experiments. In his educational work, Martin [19] also uses sound card and white noise to measure the speed of sound in a Lucite tube filled with different gases. The advantage of our PRN method against white noise excitation is that the power density is concentrated at the chosen frequencies, while in the case of the white noise it distributes apart the entire frequency range. On the other hand, no averaging is needed to realize a high quality measurement. Since the parameters of the PRN (the frequency range and the number of frequency components) are adjustable, we can choose 1 Hz refresh rate in the tuning fork experiment to obtain better frequency resolution. The block diagram of the system there is shown on figure 10. The frequency response of the tuning fork measured in the case of an acoustic excitation is seen on figure 11.

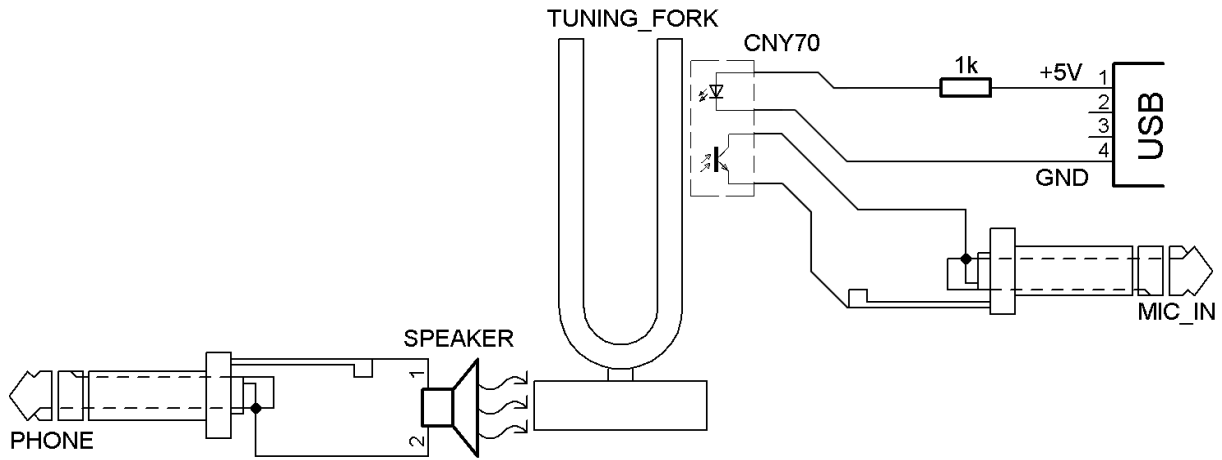

**Figure 10.** The speaker excites the resonator box of the tuning fork placed in front of it. The modulated light of the infrared LED supplied by the USB port is measured by the phototransistor connected with the microphone input [20].

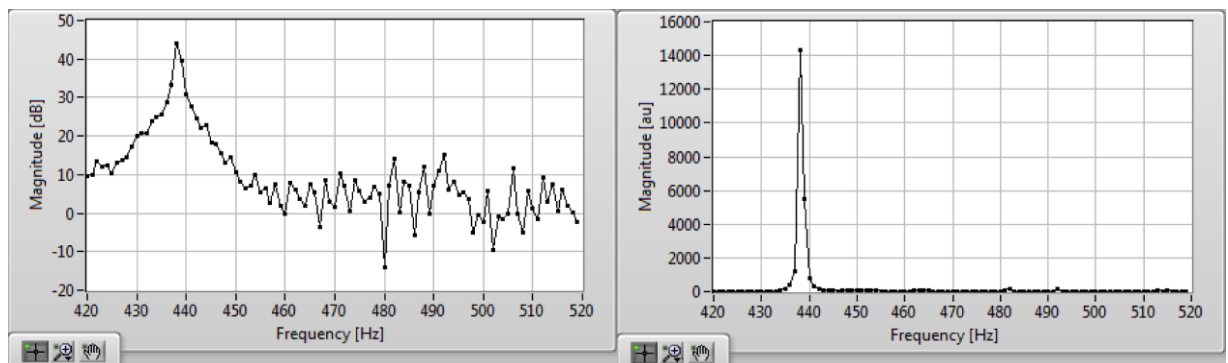

**Figure 11.** The frequency response of the tuning fork in the case of an acoustic excitation at 1 Hz frequency resolution. The spectrum of the oscillation is shown also in dB scale (left) and in linear scale (right).

One can make the tuning fork measurement more impressive by mounting different small weights to it to obtain a shift in the resonant frequency. Many other mechanical resonators can be used and resonant frequency dependence on size, shape and material can be visualized easily.

*2.2.2. Mechanical excitation with rubber hammer*

The rubber hammer excitation is similar to the impulse-like excitation using in electrical experiments. If we hit the tuning fork with a rubber hammer, we can see a long lasting damping sine wave (see the left panel of figure 12). Due to the higher mechanical energy, the signal-to-noise ratio is better than in the case of the speaker excitation, as shown on the right panel of figure 12.

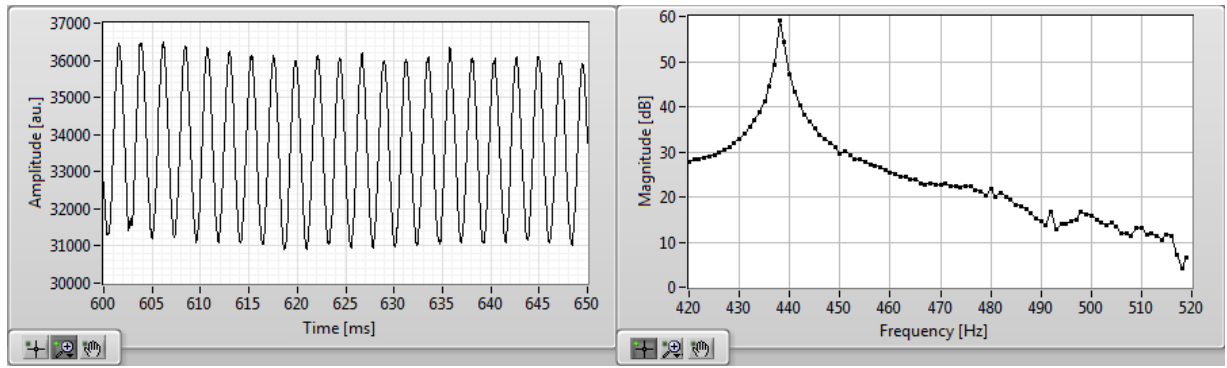

**Figure 12.** The oscillating tuning fork hit by the rubber hammer (left) and the frequency response of the tuning fork (right).

## 3. Summary

In this paper we have shown a brief description of a sound card based measuring system which was developed to study and demonstrate mechanical and electrical resonances and to measure transfer functions in real time. Our system can be very useful at the different levels of physics education.

Lecturers can use the system for experimental demonstration during oral presentations for undergraduate students. Undergraduate and graduate students can experiment in a laboratory and due to the very low cost every student can have a separate system allowing much more efficient learning. The open-source graphical user interface and graphical algorithmic language software written in the most common environment (LabVIEW) and simple hardware makes the measurement and signal processing highly transparent. Student can also benefit from learning about modern instrumentation, calibration, sensor applications, data conversion, digital signal processing and even circuit simulation. The real-time operation can help a lot to understand transfer functions, linear system analysis and behavior. It is important to note that students can easily build and use the system and they can make experiments even at home. They can also modify the software and hardware; they can be inspired to use their creativity.

We would like to emphasize that both electrical and mechanical systems can be investigated in a universal way by the use of sensors and data converters that further improves the knowledge about today's modern, computerized tools applied in household equipments, smart phones and industrial machines, robots.

Since the operation of the system covers rather wide range of physical and technical fields it is important not only in the education of students of physics and physics teaching, but also helpful for students of engineering.

Considerable proportion of physicists is employed in technical-oriented, engineering-like jobs therefore they can benefit from learning about and using similar instrumentation and technical methods, signal processing. The work of PhD students mainly of educational research, training of physics teachers and high-school demonstrations can also be effectively supported.

Our experimental set-ups are easy to rebuild and develop, and we provide a detailed documentation [15]. The software is fully open-source and it is also publicly available as an executable (.exe) file. The users only need the free LabVIEW Run-Time Engine software to use it [24]. Our software run both on Windows and Linux operation systems [24], and can be ported to various National Instruments devices as well.

**Acknowledgments**

Real-time analysis of mechanical and electrical resonances with open source sound card software 12The publication/presentation is supported by the European Union and co-funded by the European Social Fund. Project title: "Telemedicine-focused research activities on the field of Mathematics, Informatics and Medical sciences" Project number: TÁMOP-4.2.2.A-11/1/KONV-2012-0073.**References**

[1] Ronen, M. and Eliahu, E. 2000 Simulation - a bridge between theory and reality: the case of electric circuits *Journal of Computer Assisted Learning* 16(1), 14-26.

[2] Graham Wild, Geoff Swan, and Steven Hinckley 2011 Computer based experiments for off-campus teaching and learning of AC electricity *Australasian Association for Engineering Education Conference* 5-7 December 2011, Fremantle, Western Australia

[3] W. C. Magno et al. 2007 Probing a resonant circuit with a PC sound card *Am. J. Phys.*, Vol. 75, No. 2, February 2007

[4] Marion O. Hagler and David J. Mehrl 2001 A PC with Sound Card as an Audio Waveform Generator, a Two-Channel Digital Oscilloscope and a Spectrum Analyzer *IEEE Transactions on Education* vol. 44, no. 2,

[5] *A sound card oscilloscope 1*: http://www.zeitnitz.de/Christian/scope_en

[6] *A sound card oscilloscope 2:* http://www.virtins.com/

[7] *A sound card oscilloscope 3:* http://zelscope.com/download.html

[8] *A sound card oscilloscope 4:* http://www.zen22142.zen.co.uk/Prac/winscope.htm

[9] *A sound card oscilloscope 5:* http://www.brownbear.de/toc.htm

[10] *A sound card oscilloscope 6:* http://www.qsl.net/dl4yhf/spectra1.html

[11] *A sound card oscilloscope 7:* http://www.sillanumsoft.org/download.htm

[12] *A sound card oscilloscope 8:* http://www.iw5edi.com/software/oscillometerxz

[13] *A sound card oscilloscope 9:* http://www.dxzone.com/dx10805/cobracom.htm

[14] *A sound card oscilloscope 10:* http://www.fatpigdog.com/SpectrumAnalyzer

[15] http://www.noise.inf.u-szeged.hu/edudev/RealTimeAnalysisOfResonances/

[16] Datasheet of the Cmedia CM119A external soundcard http://www.hardwaresecrets.com/datasheets/CM119A.pdf

[17] A free electronic simulation software by Texas Instruments: http://www.ti.com/tool/tina-ti

[18] Steven W. Smith 1997 The Scientist & Engineer's Guide to Digital Signal Processing *California Technical Pub*. 1st edition ISBN-13: 978-0966017632, http://www.dspguide.com/

[19] Brian E. Martin 2001 Measuring the speed of sound — Variation on a familiar theme *The Physics Teacher* Volume 39, Issue 7, pp. 424

[20] Zoltán Gingl, Róbert Mingesz, Péter Makra and János Mellár 2011 Review of sound card photogates *Eur. J. Phys.* 32 905

[21] Graham Wild and Geoff Swan 2010 Acoustics Education: Experiments for Off-Campus Teaching and Learning *Proceedings of 20 th International Congress on Acoustics,* ICA 2010 23-27 August 2010, Sydney, Australia

[22] Datasheet of the applied optosensor: http://www.vishay.com/docs/83751/cny70.pdf

[23] Smilen Dimitrov 2010 Extending the soundcard for use with generic DC sensors *Conference on New Interfaces for Musical Expression (NIME 2010)*, Sydney, Australia

[24] LabVIEW Run-Time Engine for use of the written LabVIEW software http://joule.ni.com/nidu/cds/view/p/id/3433/lang/hu